# On Information Processing Limitations In Humans and Machines

*Keynote Lecture delivered as a videopresentation at the 1st International Electronic Conference on Information IECI-2021, online, December 2021.*

**Birgitta Dresp-Langley**

*Centre National de la Recherche Scientifique CNRS, ICube UMR 7357, Strasbourg University, France*

birgitta.dresp@cnrs.fr

## Abstract

Information theory is concerned with the study of transmission, processing, extraction, and utilization of information. In its most abstract form, information is conceived as a means of resolving uncertainty. Shannon and Weaver (1949) were among the first to develop a conceptual framework for information theory. One of the key assumptions of the model is that uncertainty increases linearly with the amount of complexity (in bit units) of information transmitted or generated. A whole body of data from the cognitive neurosciences has shown since that the time of human response or action increases in a similar fashion as a function of information complexity. This paper will discuss some of the implications of what is known about the limitations of human information processing for the development of reliable Artificial Intelligence. It is concluded that novel conceptual frameworks are needed to inspire future studies on this complex problem space.

## Introduction

This paper relates to Information Theory (Shannon and Weaver, 1949). Information therein is aimed at resolving uncertainty in complex systems. At this stage, computers as we know them today did not exist. The problem with this theoretical framework relates, among other, to the fact that the general information theory premise specifies neither the nature « information », nor the nature of « complexity » or « uncertainty ». The Shannon-Weaver Law stipulates that uncertainty in information systems increases linearly with the amount or complexity (in *bit* units) of information transmitted or generated (Shannon and Weaver, The Mathematical Theory of Communication, 1949; University of Illinois, Urbana III). The Hick-Hyman Law was formulated subsequently on the basis of studies in psychology showing that uncertainty in psychophysical systems is directly reflected by the time of human response to stimuli in the environment, which increases with stimulus complexity (Hick, Quarterly Journal of Experimental Psychology. 1949; 4 (4:1): 11–26; Hyman, Journal of Experimental Psychology, 1953; 45 (3): 188-196).

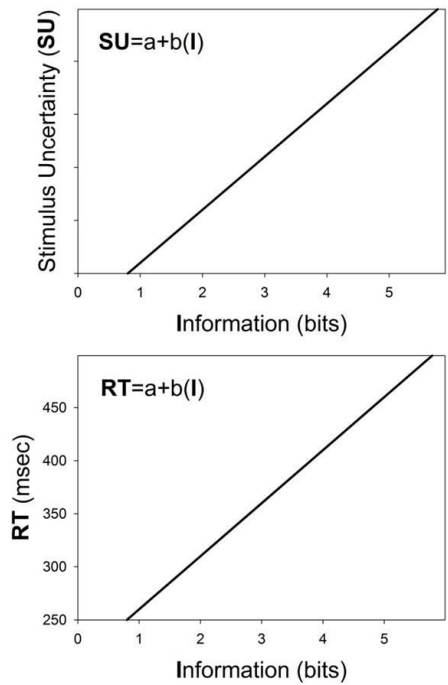

**Figure 1:** The Shannon-Weaver (top) and Hick-Hyman (bottom) Laws of information uncertainty and human response time

This leads to consider the problem space of human minds and machines under a new light, where the concepts of *information* and *uncertainty* relate to the complexity in physical or computational systems on the one hand (machines), and to the complexity of the living systems (brains/minds) that conceived and created them on the other. With the rise of Artificial Intelligence (AI) with different levels of functional complexity and autonomy and the new possibilities offered by quantum computing, we are faced with new challenges that require new paradigms for scientific investigation.

**Levels of Uncertainty in Minds and Machines**

Whether we are dealing with physical or computational systems created by human minds or with the mind as a creative system changes the both the nature and the levels of complexity to be considered.

*Sensory Uncertainty*

*Sensory Uncertainty* is a conceptual workspace that relates to the detection of signals (*stimuli*) in the environment by living sensory systems and the encoding of such signals (visual, auditory, somatosensory) by the brain. The time taken by human individuals to detect sensory signals in the environment decreases as a power function of the increase in signal intensity (Piéron, The Sensations. 1952*;* Yale University Press*)*. Figure 2 shows this type of sensory uncertainty in the case of a visual stimulus.

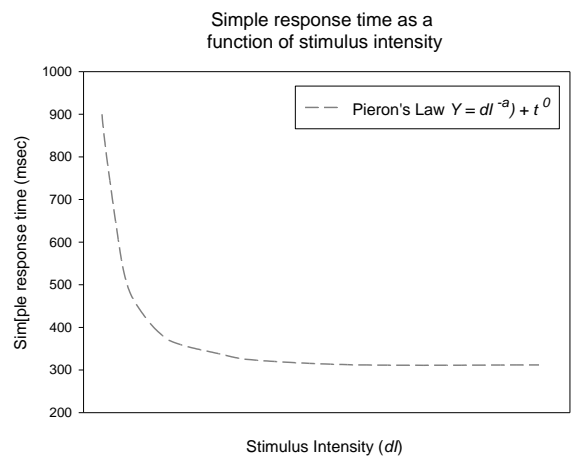

**Figure 2:** The Piéron function as a measure of sensory uncertainty in the human brain.

*Cognitive Uncertainty*

Cognitive uncertainty corresponds to system states that require regulation of representations of the environment in order to obtain better prediction and adaptation (Mushtaq, Bland, Schaefer, Uncertainty and Cognitive Control. Frontiers in Psychology, 2011;2:249). In humans, functional interaction between conscious and non-conscious cognitive workspaces enables decision making under conditions of high uncertainty *(*Dresp-Langley, Why the Brain Knows More than We Do: Non-Conscious Representations and Their Role in the Construction of Conscious Experience. Brain Sci. 2012; 2: 1-21)*.*

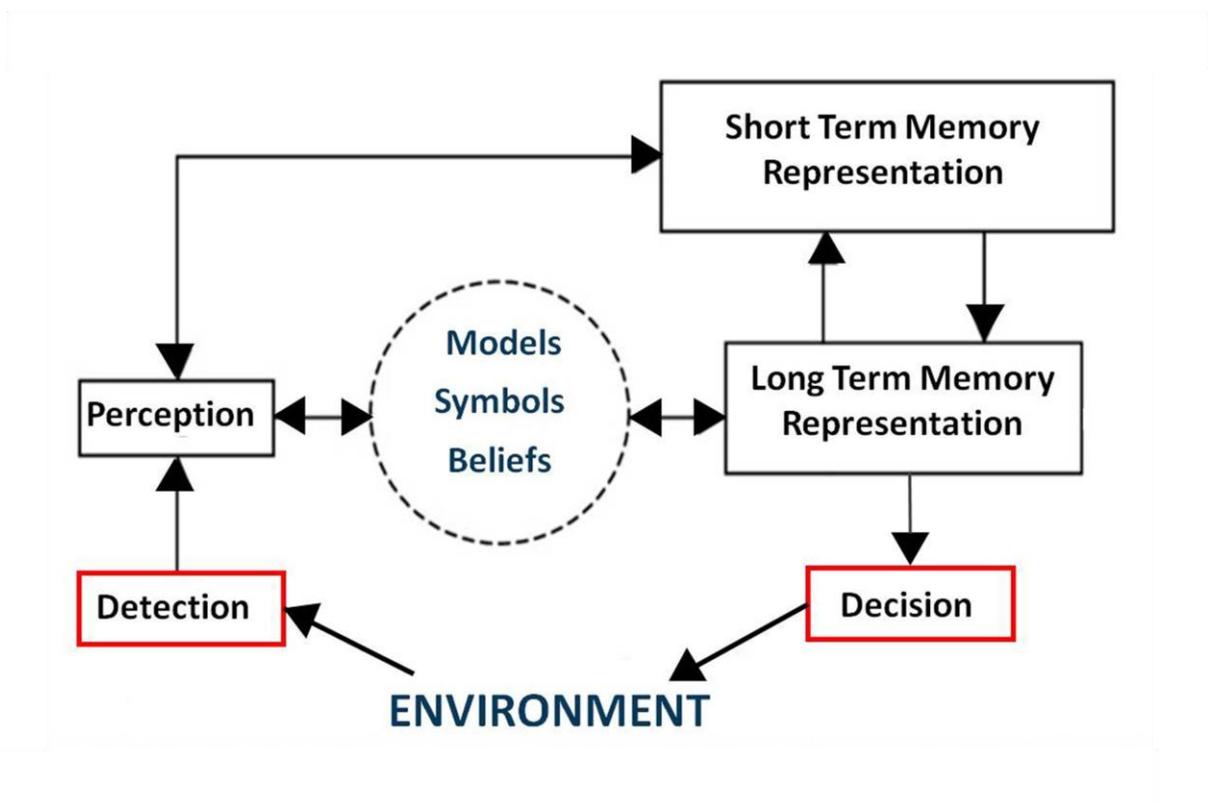

**Figure 3:** Cognitive uncertainty involves multiple levels of mental representation between perception and decision where subjectivity (beliefs) and unconscious representations interact with the conscious workspace of the human mind.

*Uncertainty in Neural Networks*

Uncertainty in neural networks is related to a variety of interdependent factors such as the complexity of the input data, the complexity of the functional neural network architecture designed to process the data, the adequacy and complexity of the learning algorithms, and the complexity of input and output dimensionalities. Some of these issues here may be illustrated by comparing properties and functional architectures of a Deep Neural Network (DNN) and a Self-Organizing Map (SOM), as suggested here below in Figure 4.

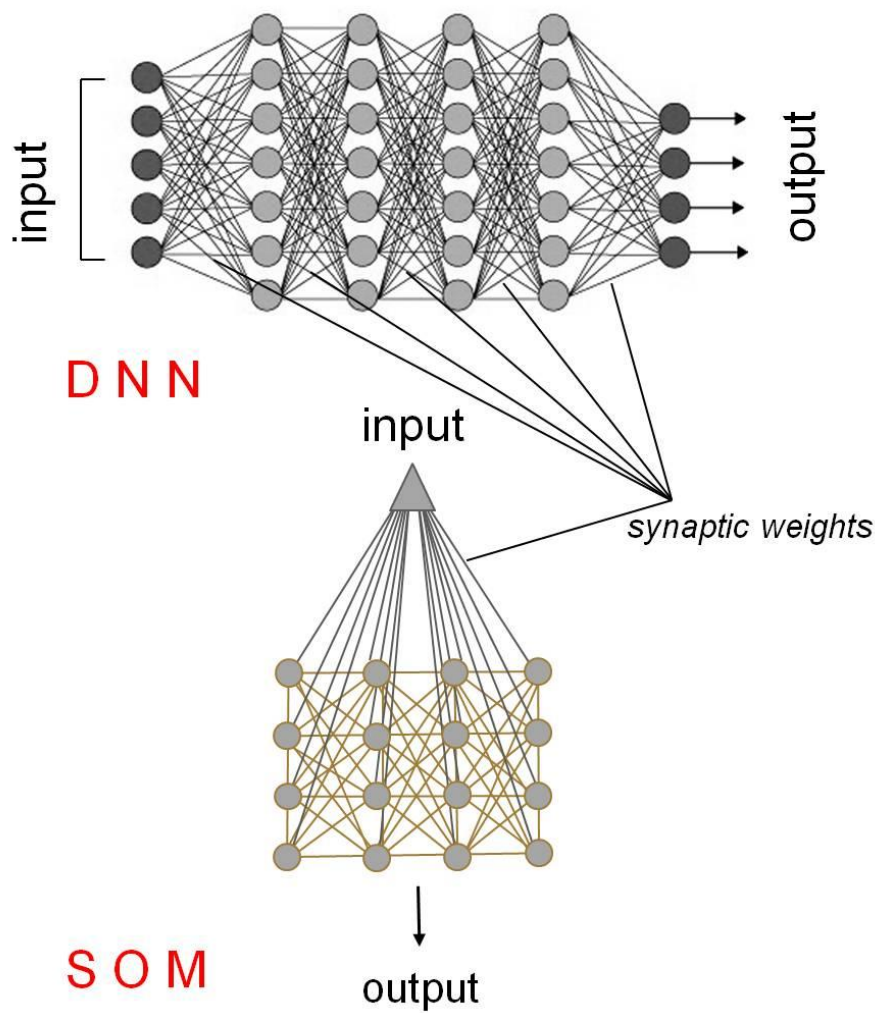

**Figure 4:** A DNN (top) has a complex network architecture with many functionally non-defined hidden layers und unknown levels of system uncertainty. A SOM (bottom) represents a distribution of input data vectors using a finite set of models. As a measure of system uncertainty, the quantization error (QE) of an input vector, expressed as the Euclidean norm of the difference between input vector and best-matching model, may be computed (Kohonen , Nieminen, Honkela, On the Quantization Error in SOM. Lecture Notes in Computer Science. 2009; 5629. Springer, Berlin, Heidelberg).

*Uncertainty In Artificial Intelligence*

To approach the concept of uncertainty in Artificial Intelligence (AI) one may consider their different levels of autonomy. Each level generates combinations of different types of human and/or machine uncertainties at the levels of signal detection, data cognition, and decision making. We may use the three-level classification for AI autonomy referred to in the domain of autonomous weapons systems (Boulanin and Verbruggen, Mapping the Development of Autonomy in Weapon Systems. 2017; The Stockholm International Peace Research Institute):

**Level 1 AI: Human controlled** - *'human on the loop'*: human agent initiates and controls all steps in the process

**Level 2 AI: Semi-autonomous** - *'human in the loop'*: human has control over some of the steps in the process

**Level 3 AI: Fully autonomous** - *'no human in the loop'*: human has no control over any step in the process

While level 1 and level 2 AI combine uncertainty in human mind and machine, level 3 AI uncertainty involves that of the machine only. The reliability of level 3 AI depends entirely on the way it has been designed (algorithms and control functions). Level and level 2 AI may benefit from the life-long expertise and intuition of human expert minds "on" or "in the loop" while level 3 AI will execute what has been programmed without any further possibility of control.

**Conclusions**

Shannon & Weaver's post-war Information Theory is challenged by contemporary cognitive neuroscience and the rise of neural network learning and AI. A novel conceptual framework for what is to be understood by « information », « complexity », and « uncertainty » needs to be carved out to develop new paradigms for research. The processing constraints and limitations of human brains and machines need to be studied in domain and application specific contexts. The conscious processing limitations of a human agent in any context are compensated for by non-conscious processes that run in massively parallel, dedicated resonant networks of the human brain. Interactions between « conscious » and « non-conscious » cognitive workspaces cannot be implemented in current AI. Under conditions of critically high uncertainty, the human expert can resort to decisions on the basis of intuition, the machine (AI) cannot.

**References**

Shannon and Weaver, The Mathematical Theory of Communication, 1949; University of Illinois, Urbana III.

Hick, Quarterly Journal of Experimental Psychology. 1949; 4 (4:1): 11–26.

Hyman, Journal of Experimental Psychology, 1953; 45 (3): 188-196).

Piéron, The Sensations. 1952; Yale University Press.

Mushtaq, Bland, Schaefer, Uncertainty and Cognitive Control. Frontiers in Psychology, 2011;2:249.

Dresp-Langley, Why the Brain Knows More than We Do: Non-Conscious Representations and Their Role in the Construction of Conscious Experience. Brain Sci. 2012; 2: 1-21.

Kohonen, Nieminen, Honkela, On the Quantization Error in SOM. Lecture Notes in Computer Science. 2009; 5629. Springer, Berlin, Heidelberg.

Boulanin and Verbruggen, Mapping the Development of Autonomy in Weapon Systems. 2017; The Stockholm International Peace Research Institute.